# Interaction Mechanism and Response of Tidal Effect on the Shallow Geology of Europa

*Yifei Wang* [1]

*Beijing University of Aeronautics and Astronautics*
*Department of Materials Science and Engineering*
*Undergraduate*

## ABSTRACT

Europa has been confirmed to have a multilayered structure and complex geological condition in last two decades whose detail and cause of formation remains unclear. In this work, we start from analyzing the mechanism of tidal effect on satellite's surface and discuss if interaction like tidal locking and orbital resonance play an important role in tidal effect including heating, underground convection and eruption. During discussion we propose the main factors affecting Europa's tidal heat and the formation mechanism of typical shallow geological features, and provide theoretical support for further exploration.

*Keywords:* Europa, tidal heating, tidal locking, orbital resonance

## 1. INTRODUCTION

According to existing observations by *Galileo* radio-tracking/gravity data and *Galileo* SSI images, Europa has a complex structure and distribution of different components including solid ice layer, warm liquid ocean, local convection and eruption between solid and liquid region, and hard brittle lithosphere. And for Jupiter's three moon system--IO, Europa and Ganymede, these are in the rotation and revolution of synchronous pattern which called tidal locking. Besides, it is confirmed that IO, Jupiter and Ganymede are in the situation of orbital resonance, whose orbital period ratio is about 1:2:4 and the ratio remains constant with time.

Their effect on the surface is periodic during revolution. From Ogilvie and Lin's work[2], a response to tidal force is separated in two parts: an equilibrium tide and a dynamic tide which is irrelevant with time and the angular position to Jupiter and other satellites because of periodic interaction to the subsurface structure. However, there is no recognized and unified explanation for the geological structure, composition and formation for Europa so far. And For shallow landscape especially the large area of glacial rift valley, a convincing explanation is missing. In this paper, we propose that tidal effects which mainly includes deformation in surface and inner structures and tidal heating, are mainly caused by the process where Europa tends to be tidal locked, and the gravitational torque is the motivation, and orbital resonance, or Laplace among 3 satellites, minimizes tidal thermal effects. We also propose a reasonable forming process which may provide theoretical support and predictions for further detection.

## 2. DYNAMIC ANALYSIS AND MODELING

### 2.1. *Tidal Deformation on Europa's Surface*

For the angle between Jupiter's equatorial plane and Europa's orbital plane is $0.470°$. And it's easy to prove that the vector diameter $\boldsymbol{r}$ of Europa satisfies

$$A\ddot{\boldsymbol{r}} + B\dot{\boldsymbol{r}} + C\boldsymbol{r} = \boldsymbol{0}$$

A, B and C are constants which are irrelevant with the position vector, $\ddot{\boldsymbol{r}}$, $\dot{\boldsymbol{r}}$ and $\boldsymbol{r}$ are in the same certain plane, accordingly, below we will treat it as the approximation of coplanar.

To simplify the analysis, we consider system which only includes a central celestial body and a single satellite. Since the size of Jupiter is much larger than Europa and the deformation of tidal effect on Jupiter is much larger than that on Europa, the satellite was considered as SNREI body which is spherically symmetric, nonrotational and elastic isotropic, as shown in **figure 1**.

---

[1] Contact Information: yifeiwangmse2017@buaa.edu.cn Address: 37 Xueyuan Road, Haidian District, Beijing, China

[2] See in REFERENCES [3]

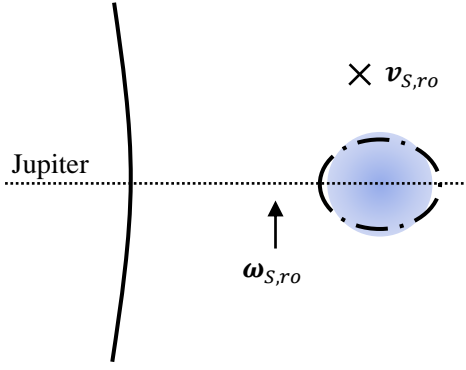

**Fig. 1**

*Schematic diagram of tidal phenomena*

Suppose Europa's mass is $m$ and the central celestial body's mass is $M$, Europa's radius $R$, and the gravitational constant is $G$. The radius from any point on Europa to the center of mass of Jupiter is $\boldsymbol{r}$ with unit vector $\boldsymbol{e_r}$. Take the center of Europa as the reference frame, the inertial forces at all points $\boldsymbol{r}$ reverse from the vector diameter, which is

$$d\boldsymbol{F_0} = -\frac{GmdM}{r^2}\boldsymbol{e_r}$$

Therefore, microbody $dM$ is taken from the nearest and farthest points of Europa from the center of Jupiter, and the force received by the two places, namely the tide-generating force, was respectively

$$d\boldsymbol{F_1} = GmdM\left(\frac{1}{(r-R)^2} - \frac{1}{r^2}\right)\boldsymbol{e_r}$$

$$d\boldsymbol{F_2} = GmdM\left(\frac{1}{(r+R)^2} - \frac{1}{r^2}\right)\boldsymbol{e_r}$$

Therefore, different distances from Europa cause the different gravitational attraction to microbody. And the parts facing and departing Europa have opposite directions of tide-inducing force, which makes the surface be stretched. Qualitatively, the integral result must be $\boldsymbol{F_1} - \boldsymbol{F_2} > 0$, which causes deformation and reset of the surface shape and elevation.

If Europa is only attracted by Jupiter's gravity, then due to the fact that $r > 400R$, we have $r^2 \gg R^2$. The sum of two tide-generating forces

$$\Delta F = \frac{GmM_{eq}}{r_{eq1}^2} - \frac{GmM_{eq}}{r_{eq2}^2} \approx \frac{4GmM_{eq}}{\frac{r}{R}\left(r^2 - 2R^2 + \frac{R^4}{r^2}\right)}$$

$$= \frac{4GmM_{eq}R}{r^3}$$

while $M_{eq}$ is the equivalent mass acted on by Europa, and $r_{eq1}$ and $r_{eq2}$ are the equivalent distance acted on the front and rear surfaces respectively. These equivalent quantities have the same dimension with corresponding quantities, and in positive correlation with them, but can't be precisely solved due to practical geometrical shape and situation. For two bodies with constant mass and volume, $M_{eq}$ is a constant value. And for Europa, $r_{eq}$ is approximately equal to the distance from Europa's surface to the center of Jupiter.

Investigate the surface strain due to tidal force. Suppose the magnitude of unilateral deformation is $\Delta r$ when it reaches stability, and the equivalent tensile area of Jupiter is $A_{eq}$, where $A_{eq}$ is the sum effect of external tension and vertical internal pressure on both sides. This rationality is due to the fact that the vertical internal pressure effect is linearly proportional to the tensile effect on both sides. According to dimensional relation we have $A_{eq} \propto R^2$. Suppose the young's modulus at any given position on the surface of Europa to be $E_S$, there is

$$R、E_S、A_{eq} = const.$$

So we can get the stress-strain relationship which is

$$\Delta F = E_S \frac{\Delta r}{R} A_{eq} \propto \Delta r$$

It can be seen that the $\Delta F$-$\Delta r$ presents a linear relationship, and the size of $\Delta F$ directly determines the degree of Europa's tidal effect. Substitute this formula into the above formula to define the tidal strain $\varepsilon$ whose size represents the deformation effect, so we get

$$\varepsilon = \frac{\Delta r}{R} = \frac{4RGmM_{eq}}{E_S A_{eq}}\frac{1}{r^3} \propto \frac{1}{r^3}$$

So $\Delta r$-$r^3$ is inversely proportional. In conclusion, the 1. larger mass and radius 2. the smaller distance 3. the larger young's modulus of the tidal body: the more significant the deformation effect tidal interaction can cause. Due to the complexity of Europa's geological composition and structure Different areas of equivalent young's modulus and density will be the first motivation of different deformation conditions, causing a lot of collisions and a series of physical changes in shallow geology.

2.2. *The Evolution of Tidal Locking*

The tidal effect on Jupiter works in exactly the same way on the Earth caused by the difference of Moon's gravity. Tidal effect will make the system gradually close to the tidal locking state. When a satellite has a tidal effect, its surface will deform to a certain extent, and the celestial body can no longer be regarded as a sphere. From this analysis below we will see the reason why the tidal phenomenon leads the system to approach tidal locking, and this process is the second motivation of geological interaction and generation of tidal heating.

We first define the system's initial state as: Europa is captured by Jupiter and has just formed a stable orbit. When tidal interaction occurs, Europa's body is approximately assumed to be an ellipsoid, and the direction of its long axis is the line between the center of Jupiter and Europa. Due to the tidal phenomenon, the surface of the earth is deformed. Correspondingly, the surface which is perpendicular to both sides of the long axis will fall or concave.

For most celestial bodies, the direction of rotation is the same as the direction of revolution, which is only discussed later. When Europa was initially captured, the angular velocity was satisfied $\omega_{S,re} \neq \omega_{S,ro}$.

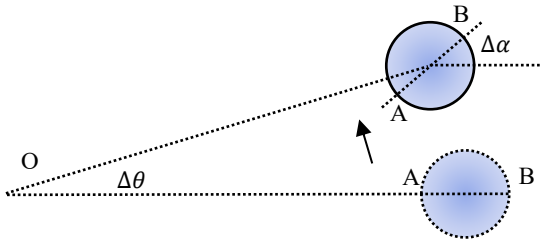

*Fig. 2 Schematic diagram of tidal locking*

According to **figure 2**, investigate the two states of the single satellite model. Europa reaches position 2 after rotating from position 1 for a small period of time $\Delta t$. In the process from position 1 to 2, the micro precession angle of Europa is $d\theta$, and $d\alpha$ is the angle of rotation. It's obvious that

$$d\alpha - d\theta = (\omega_{S,re} - \omega_{S,ro})dt = \Delta\omega dt$$

Tidal energy Q can be generated due to the interaction between the surface of the earth (such as squeezing, tribological heating, etc.). After a limited period of time, there are $d\alpha - d\theta = 0(t = t_l)$ and $\frac{d\alpha}{dt} = \frac{d\theta}{dt} = 0(t \geq t_l)$, which means tidal locking state is approached and $\omega_{S,ro} = \omega_{S,re}(t \geq t_l)$. Since Europa's radius is much smaller than Jupiter's, it can be considered that the projection of the force exerted by mass microbodies at A and B on the connection direction of the center of the two bodies is equal, and the difference mainly lies in the projection size perpendicular to the connection direction. Due to geometrical relation, the small angle between $\overline{AB}$ and the line satisfy

$$d\theta_{AB} = \frac{\overline{AB}}{2}\Delta\omega dt/r$$

So the vertical components of force at A and B are

$$dF_{Ay} = dF_A \cdot \sin d\theta_{AB} = dF \cdot \overline{AB}\Delta\omega dt/2r$$

$$dF_{By} = dF_B \cdot \sin d\theta_{AB} = dF \cdot \overline{AB}\Delta\omega dt/2r$$

They have equal value and opposite direction, which is equivalent to a gravitational torque. The differential form is

$$d\boldsymbol{M}_G(\theta = 0) = -dF \cdot \overline{AB}^2 \Delta\omega \cdot dt/2r \cdot \boldsymbol{e}_\omega$$

while $\boldsymbol{e}_\omega$ is a unit vector in the same direction as the net angular velocity, and only for a pair of special points (A, B).

Investigate the state of mass point pairs around satellite's surface. The specified point pair (A, B)'s angular position $(\varphi_A, \varphi_B) = 0$ and the angle increases with respect to counterclockwise. It can be found that all the point pairs' positions where the gravitational torque increases and prevents the rotation are $(\varphi_i, \varphi_j) \in$

$\left(0, \frac{\pi}{2}\right)$ while $\boldsymbol{M}_G$ increases and reverses $\boldsymbol{e}_\omega$

$\left(\frac{\pi}{2}, \pi\right)$ while $\boldsymbol{M}_G$ decreases and aligns with $\boldsymbol{e}_\omega$

For $\left(0, \frac{\pi}{2}\right)$, combined with the above equation, we can get

$$d\boldsymbol{M}_G(\theta, t) = -\frac{\overline{AB}^2 \Delta\omega dt}{2r} dF \cos\theta \, \boldsymbol{e}_\omega$$

Therefore, in a certain angular range，the gravitational torque will reduce $|\Delta\boldsymbol{\omega}|$ and occurs an irreversible process of tidal heat generation, and finally tends to be stable where $|\Delta\boldsymbol{\omega}| = 0$, proves the inevitability of tidal locking under certain conditions. Besides, we can see that tidal effect is a periodic interaction and is related with fluid's oscillation in chapter **3**, where tidal heating's diffusion and decay play a decisive role.

2.3. *Orbital Resonance's Impact on Tidal Locking*

Since Europa is not only tidally locked by Jupiter but also has a Laplace resonance with IO and Ganymede, which also happen to be tidally locked by Jupiter, due to this coincidence, we suspect that orbital resonance plays a role in the tidal locking process of the celestial body. The cause and deformation of tidal phenomena can be analyzed and explained from the perspective of force, but it would be extremely difficult to analyze and solve the force problem of multi-body resonance. Therefore, we discuss the role of orbital resonance by analyzing the process of orbital evolution and the variation and transformation of quantities such as tidal heating.

Any satellite shall go through the following three processes from the initial state to orbital resonance (as shown in *Fig.* 3):

**1.** Get captured from free motion and enter the orbit.
**2.** Make self-coordination to achieve tidal locking.
**3.** Achieve orbital resonance by long-term gravity by other satellites which cannot be ignored.

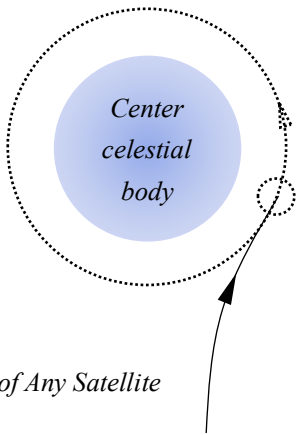

*Fig. 3*
*The Capture of Any Satellite*

The orbit radius and orbital speed of a satellite are determined by its initial kinetic energy when it was captured by Jupiter. Since multi-body problem can't be analyzed and the interaction between Jupiter and its satellites is much larger than that between satellites, the interaction between the three satellites was ignored in process **1**, and mechanical energy is considered to be composed only of the gravitational potential energy and its own kinetic energy brought by Jupiter.

For Europa, the blocking of gravitational torque on rotational kinetic energy is much smaller than that of gravitational force on translational energy, so the variation of rotational energy is ignored in this section.

For any satellite's process in beginning to form in orbit, being captured by Jupiter's gravity and achieving stable orbits, apparently these gravitations from other satellites can be neglected, and Jupiter's mechanical energy in satellites' gravitational field can be considered as constant, therefore, satellite in process **1** can be considered to satisfy the conservation of the sum of mechanical energy and thermal energy.

Might as well set $r_\infty$ up for infinitely long time after the ideal stable orbits, and set the initial capture position $r_0$ as the zero-potential point. In later discussion, the selection of $r_0$ must satisfy certain conditions. So we have

$$E_\infty = \left[E_{p0} - \int_{r_0}^{r_\infty} F(r)dr\right] + \frac{1}{2}mv_\infty^2$$

with $F(r) = -\frac{GMm}{r^2}$ and $m\frac{v_\infty^2}{r_\infty^2} = \frac{GMm}{r_\infty^2}$. So we get

$$E_\infty = (\frac{1}{r_0} - \frac{1}{2r_\infty})GMm$$

Due to the conservation of satellite's energy which is assumed above, $E_\infty = E_0 = E_{p0} + \frac{1}{2}mv_0^2 = \frac{1}{2}mv_0^2$, we can get the steady-state orbit radius after single satellite's capture which is

$$r_\infty = \frac{1}{\frac{2}{r_0} - \frac{v_0^2}{GM}}$$

We can see that $r_\infty$ is only related to the mass of Jupiter, $r_0$ and initial capture velocity rather than mass of itself. And notice that $\frac{v_0^2}{GM} > 0$, there must be

$$1 < \frac{r_0}{r_\infty} < 2$$

It's no doubt that the initial capture position $r_0's$ selection is important. Based on previous discussion, it should be defined as the maximum boundary radius where gravitational effect from Jupiter cannot be ignored.

This can explain IO, Europa and Ganymede's relative angular positions are very close to $0:\pi:0$ (as shown in ***Fig. 4***). The angular velocity of steady-state satellite is mainly related to its orbital radius, followed by the position of the surrounding satellite. The difference in angular velocity between perihelion and aphelion is very small, and the perturbation of the three satellites can be ignored. Therefore, its angular velocity of revolution only depends on the orbital radius.

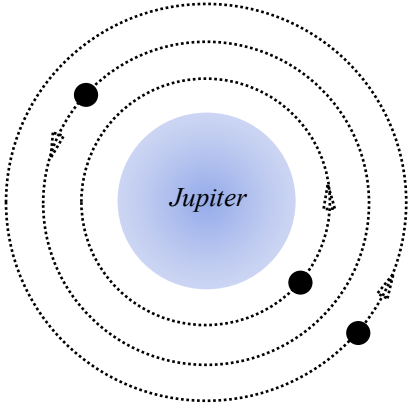

***Fig. 4*** *Orbital Resonance Positions of IO, Europa and Ganymede*

When the moon has not entered the tidal locking state, it will slow down and rotate under the action of the moment of resistance according to 2.2. The evolution of the single-moon and Jupiter system follows

$$L_0 = \sum L_i$$

$$E(r) = \sum E_{ki} - \frac{GMm}{r}$$

while $\sum L_i = J_{P,ro}\omega_{P,ro}(r) + J_{S,ro}\omega_{S,ro}(r) + J_{S,re}\omega_{S,re}(r)$ and $\sum E_{ki} = \frac{1}{2}[J_{P,ro}\omega_{P,ro}^2(r) + J_{S,ro}\omega_{S,ro}^2(r) + J_{S,re}\omega_{S,re}^2(r)]$ ($r$ is orbit radius of the satellite), which are respectively the angular momentum and kinetic energy of the system. As the satellite rotation slows down, $r$ increase and system's mechanical energy will decrease. In other words, during the satellite deceleration, tidal heat is generated due to the friction between surface and internal matter, which is

$$\delta Q = E(r) - E(r + dr)$$

Boundary conditions for equations above are

$$\omega_{S,ro}(r_l) = \omega_{S,re}(r_l)$$
$$Q_l = E(r_l) - E(r_0)$$

While subscript $l$ refers to the locking state in corresponding quantities. $r_0$ indicates the initial spacing where satellite is just captured by the Jupiter.

Then we investigate the orbital evolution process of multiple satellites. Since the interaction among satellites is much smaller than that between satellite and Jupiter, it can be considered that the joint action of IO, Europa and Ganymede on Jupiter is approximately three moons act separately on the linear superposition, which avoid the impossible analysis about multi-body interaction. Same as above, we have

$$L_S = \sum_{i=1}^{3}[J_{Si,ro}\omega_{Si,ro}(r) + J_{Si,re}\omega_{Si,re}(r)]$$

$$L_S + L_P = L_0$$

$$E_{kS} = \frac{1}{2}\sum_{i=1}^{3}[J_{Si,ro}\omega_{Si,ro}^2(r) + J_{Si,re}\omega_{Si,re}^2(r)]$$

$$E_{kP} + E_{kS} - \sum_{i=1}^{3}\frac{GMm_i}{r_i} = E(r)$$

while $L_P = J_{P,ro}\omega_{P,ro}(r)$ $E_{kP} = \frac{1}{2}J_{P,ro}\omega_{P,ro}^2(r)$.

Notice that the periodic interaction of orbital resonance is much longer than the locking process. The satellite has been tidally locked until reaching Laplace resonance, for $i = 1,2,3$ and $r = r_l$,

$$\omega_{Si,ro}(r) = \omega_{Si,re}(r_l)$$

When it reaches Laplace resonance, we have

$$\omega_{S1,re}:\omega_{S2,re}:\omega_{S3,re} = A_1:A_2:A_3 = 4:2:1$$

The above is the conservation equations followed by the state variables in evolution process, and the Let the initial orientation of satellites be $\theta_{0i}(i=1,2,3)$, according to existing observations, their angular positions are respectively

$$\theta_1 = \theta_{01} + \omega_1 t = 2\pi t/T_1)$$
$$\theta_2 = \theta_{02} + \omega_2 t = \pi(1 + t/T_1)$$
$$\theta_3 = \theta_{03} + \omega_3 t = \pi t/2T_1$$

Assuming three satellites can reach the same side of Jupiter and collinear, there must be $k_1, k_2 \in Z$ make $\theta_1 - \theta_2 = 2\pi k_1$, $\theta_2 - \theta_3 = 2\pi k_2$. Get rid of $t$ and we can get

$$2k_1 - 4k_2 + 3 = 0$$

no integer solutions for $k_1$, $k_2$, contradiction, so the assumption is wrong.

1:2:4 for IO, Europa and Ganymede system therefore, is unlikely to be the location of the ipsilateral collinear, and qualitatively, the state of the ipsilateral collinear will result in the interaction between satellite is the strongest, before and after reach the state of the ipsilateral collinear, its changes such as torque, the relative potential energy, also is the strongest, the overall is not conducive to long-term stability.

When any satellite has yet to reach locking state, during the process, go through the rest of other two satellites and multiple disturbance of Jupiter, mainly reflected in the torque's value and direction and rapid changes in potential energy. Each with a satellite formation tidal locking, its gravitational torque effect becomes 0, at the same time reduces the frequency of the disturbance of the rest of the satellite. In a long period of a large number of periodic interactions, the formation of such a stable resonance state.

## 3. SHALLOW GEOLIGICAL RESPONSE TO TIDAL ACTION

Based on Galileo and its SSI observations and earlier voyager thermal simulations, Europa has four major geological features: **1.** Highly differentiated geological crust, including metal/metal sulfide core, rocky mantle, outer (near surface) water layer or 80-170km deep ocean, surface ice. **2.** Ductile deformation exists in the inner part, the lithosphere transversal migration rate is in the tens of kilometers, and local melting began in modern times. **3.** Geologically active: surface craters retain an age of 10-100 million years. The surface craters are estimated to have remained for 30 million years (10 kilometers in diameter) and are found to be consistent with the 10 kilometers of ocean ice. **4.** Has vast glacial rift valleys on surface which stretches for hundreds of kilometers and local heat flows may also exist.

### 3.1. *Geological Processes from Tidal Effect*

The discovery of a large amount of liquid water under Europa and other explorations are still in an imperfect stage and lack of explanation of the creation of the shallow geological structure. Currently, there are four main international conjectures (as shown in *Fig. 5*):

**I.** A thin, brittle, electrically conductive crust of ice covering a deep global ocean.
**II.** The crust layer of almost completely solid water, consisting of the thin, brittle ice lithosphere in the upper part and the warm, convective ice asthenosphere in the lower part.
**III.** between **I** and **II**: A thin layer of ocean, global or scattered, beneath thick, convective ice and brittle lithosphere
**IV.** between **I** and **II**: Outer layer thick convection ice crust, liquid ocean in the middle, surface fragile lithosphere beneath.

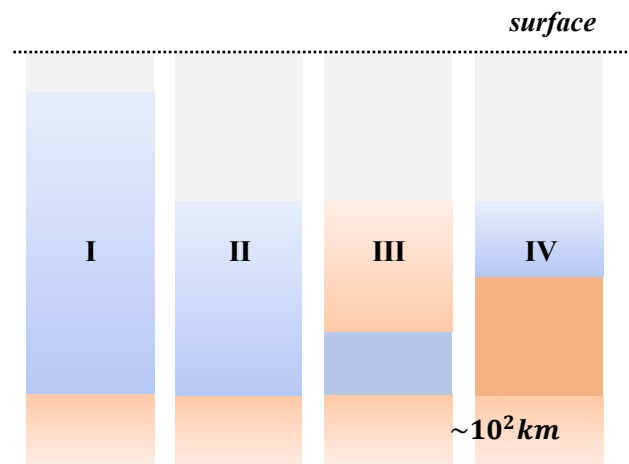

*Fig. 5 Four Main Conjectures of Europa's Shallow Geology*

According to analysis in chapter **2**, the tidal strain of the surface of the celestial body affected by tide is

$$\varepsilon = \frac{4RGMm_{eq}}{E_S A_{eq}} \frac{1}{r^3}$$

If the crystal in the geological layer is a perfect crystal of the same material with high symmetry and no defects, the deformation of tidal effect on geological layer is exactly the same at the same thickness from the center of the celestial body. Therefore, for tidal thermal effect, the heat generation power of deep geological layer is lower than that of shallow geological layer. But because Europa's mean radius $\overline{R_2} = 0.245$Ea. $= 1561$km, for shallow geology 100km about from the surface, the difference of tidal strain between different thickness is less than 10%. Therefore, depth is not the main cause of tidal effects in shallow geology.

The observation results suggest that there are local violent processes such as eruption and explosion of frozen liquid or mud, as well as a large area of glacial rift valley on the surface. We suggest two possible formation mechanism:

**I.** Occurs in areas where geologic layer friction or plate action is most intense. In the process of tidal locking, due to different angular positions and depths of the geological element by the magnitude and direction of gravity is different, that is, tidal action, combined with its own geological heterogeneity, will lead to the local first crack and slip layer, friction and collision, and rapid increase and expansion. Due to deep solid matter is relatively shallow liquidity is poorer, tide heat by convection and heat conduction, gradually accumulation will lead to local warming rapidly, so that the solid ice crystal melting or amorphous impurities into the viscous flow, increase liquidity, heat conduction more quickly to the surrounding water ice, not melt local overheating will cause explosion. On the other hand, as the modulus E decreases due to the phase transition, the strain variable of tidal effect increases, further increasing the relative displacement and friction heat generation between liquids. This positive feedback effect makes the concentrated distribution of tidal heat quickly change to uniform distribution after the internal first reaches the melting point at a certain position, and thus forms a large area of liquid ocean.

**II.** Occurs in the region with the highest fracture brittleness. The purity, crystallinity and grain state of the ice layer are different in different positions. Under the action of local stress caused by tidal forces, cracks occur in high-risk places and heat up rapidly with relative slip and friction. Unlike **I**, the local mechanical stress may be so large that it precedes the thermal concentration, causing the fluid to erupt violently. The weight of transgranular and intergranular fractures on the surface of the earth is different due to the difference of the size of the crystal, the orientation of the crystal plane and the energy at the grain boundary.

What the main conjectures have in common are a hard, brittle surface, electrically conductive ice, and a softer, convective interior, such as a soft ice sheet or a liquid ocean. Varying degrees of internal softening may be due to tidal heat. For example, during the formation of liquid ocean, tidal heat power tends to be tidally locked with Europa, and the total tidal heat production tends to an upper limit with time evolution, which determines the proportion of liquid ocean in the total shallow volume. When it is stable, it forms a state of overall orderly stratification and local chaos and turbulence, where surface glaciers and shallow layers have liquid ocean or soft ice, where dynamic equilibrium convection and rapid and massive renewal exist at the junction of the two.

3.2 *Geological Action After Reaching Resonance*

In locking and Laplace resonance state, Jupiter has no gravitational torque on the three moons, but the weak gravitational torque between the three moons still exists. Due to the action of Jupiter is much stronger than that of other satellites, it is very difficult to change the kinetic energy caused by the gravitational coupling moment between other satellites. Most of the potential energy is converted into tidal heat instead of rotational kinetic energy. Therefore, after reaching the orbital resonance state, the tidal thermal power will decrease rapidly, but will not stop. Due to the constant period ratio, the three satellites will maintain the resonance state and change their orbit radius synchronously, and the effects such as tidal heat will gradually approach 0, forming a stable geological structure and activity form.

## 4. SUMMARY AND CONCLUTIONS

Above all, we work mainly lies in: Europa and Jupiter interaction system through reasonable assumptions and approximation, toy model analyses the dynamic mechanism of the tides, tidal locking of the forming process, and get the result that the gravitational torque is a tidal locking the direct cause of formation, modeling of Jupiter and its moons system are analyzed theoretically and analyzes the evolution process of Laplace resonance characteristics and contact, etc. Will affect the state of tidal locking orbital resonance, and tending to tidal locking this process, it is such as tidal fever, the causes of surface deformation and a series of tidal effect. The explanation of the surface glacier and the shallow layer activity is put forward. On the other hand, the shortcomings in our exploration process mainly include: the actual situation of the interaction of multiple celestial bodies is too complex to be accurately modeled and the analytic solution can be obtained. The author hopes to improve the calculation and other technical details at the same time, the corresponding conjecture can be more general and reasonable. In addition, the weak effects of orbit perturbation and longitude libration of Europa are ignored in the analysis process, which need to be further improved. At the same time, the discussion of geological activities, such as internal periodic vibration and the role of molten fluid, still has a lot of room for innovation.